# Topological Semimetal KAlGe with Novel Electronic Instability

Toshiya Ikenobe[1, *], Takahiro Yamada[2], Jun-ichi Yamaura[1], Tamio Oguchi[3], Ryutaro Okuma[1], Daigorou Hirai[4], Hajime Sagayama[5], Yoshihiko Okamoto[1], and Zenji Hiroi[1]

[1] Institute for Solid State Physics, University of Tokyo, Kashiwa, Chiba 277-8581, Japan

[2] Institute of Multidisciplinary Research for Advanced Materials, Tohoku University, Sendai, Miyagi 980-8577, Japan

[3] Center for Spintronics Research Network, Osaka University, Toyonaka, Osaka 560-8531, Japan

[4] Department of Applied Physics, Nagoya University, Nagoya, Aichi 464–8603, Japan

[5] Institute of Materials Structure Science, High Energy Accelerator Research Organization, Tsukuba, Ibaraki 305-0801, Japan

**ABSTRACT:** Compounds with the anti-PbFCl structure exhibit a variety of electronic instabilities and intriguing physical properties. NaAlSi and NaAlGe are similar topological nodal-line semimetals, but they have distinct properties. NaAlSi is a superconductor at 6.8 K, whereas NaAlGe is an insulator with a pseudogap of approximately 100 K. Using the potassium–indium flux method, we succeeded in synthesizing a single crystal of KAlGe, a new anti-PbFCl compound. First-principles electronic structure calculations reveal that KAlGe is isoelectronic with NaAlSi and NaAlGe. KAlGe undergoes a metal-to-metal transition at 89 K and exhibits no superconductivity above 1.8 K. The low-temperature phase has significantly lower carrier density and extremely high mobility, similar to Dirac electron systems. Furthermore, X-ray diffraction experiments show a structural change that breaks the fourfold symmetry during the phase transition. Electron–phonon interactions may be responsible for superconductivity in NaAlSi, whereas excitonic electron–hole interactions are thought to play an important role in KAlGe and possibly NaAlGe. Our findings demonstrate that fascinating physics lies within the compound family.

## 1. INTRODUCTION

A family of compounds with intriguing structure–property relationship frequently possess a particular crystal structure.[1, 2] Examples include perovskite structures known for dielectric, magnetic, superconducting, and solar cell materials[3-7]; pyrochlore structures exhibiting magnetic frustration, spin liquids, and spin–orbit coupled metals[8, 9]; and $ThCr_2Si_2$-type structures with heavy fermion superconductivity, magnetocaloric effects, and skyrmion.[10-12]

The PbFCl-type structure (space group $P4/nmm$) and its antistructure, which exchanges cation and anion sites, are not an exception. Figure 1(a) depicts the two structures of the typical compound ZrSiS and the present compound KAlGe, respectively. Two-dimensional metallic conduction occurs in Si (Al) square nets sandwiched between Zr (Ge) sheets that are effectively separated by double S (K) sheets. Materials with these crystal structures exhibit a wide range of physical properties as a result of various electronic instabilities. ZrSiS is a typical nodal-line semimetal that belongs to the topological quantum materials.[13] In contrast to the Dirac and Weyl semimetals, which have single and double crossing points between linearly-dispersive band branches, respectively, the nodal-line semimetal has connected band crossing points that form a loop in the momentum space. Because the Dirac nodal line is located near the Fermi energy $E_F$,[14, 15] ZrSiS shows unusual electronic properties such as extremely high mobility and large magnetoresistance,[16, 17] as observed in other nodal-line semimetals such as $WTe_2$, CaAgP, and $\beta$-$ReO_2$.[18-20] More intriguing properties have been predicted, including excitonic transition and $d$-wave superconductivity in the presence of electron correlation.[21, 22] The excitonic phase transition, which is caused by electron–hole pairing and results in pseudogap opening, is particularly unusual but yet to be discovered.

Typical compounds with the anti-PbFCl structure are found in the family of iron-based superconductors. LiFeAs and LiFeP in the family exhibit superconducting transitions at 14 K and 6 K, respectively.[23-25] NaFeAs shows a nematic transition in which the electronic system's in-plane symmetry is spontaneously broken from fourfold to twofold.[26] This transition occurs to remove the degeneracy of Fe $3d_{yz}$ and $3d_{zx}$ bands. A similar nematic instability was found in α-$FeSe_{1-x}S_x$ superconductors.[27] The effect of orbital fluctuations to the superconducting mechanism has been discussed.[27, 28]

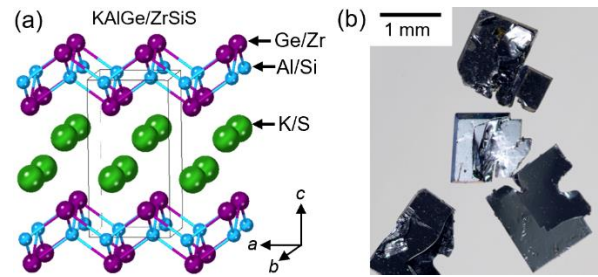

**Figure 1.** (a) PbFCl-type crystal structure of ZrSiS is shown on the left, with the antistructure of KAlGe on the right. The tetragonal quasi-two-dimensional structure is formed by stacking Si–Zr (Al–Ge) layers, which are responsible for electrical conduction, between double S (K) sheets. (b) Single crystals of KAlGe. A 1 mm square plate-shaped crystal has its primary surface as the (0 0 1) plane.

We focus on NaAlSi and NaAlGe as compounds with anti-PbFCl-type structures.[29] Despite sharing similar electronic structures,[30-33] they have completely different ground states: NaAlSi has a metallic state with a superconducting transition at 6.8 K,[34-38] whereas NaAlGe forms a pseudogap around 100 K and exhibits semiconductive resistance,[39] and is superconducting at 2.8 K after hole-doping.[40] These compounds have similar two-dimensional semimetallic states due to the intersection of Al-$s$ orbital electron bands and Si-$p$ or Ge-$p$ orbital hole bands.[41] Because the observed differences in physical properties between the two compounds are difficult to explain with similar band structures, some electronic instability that was overlooked in electronic calculations is thought to be essential. It has been pointed out that electron–electron paring via electron–phonon interactions is predominant in NaAlSi, resulting in superconductivity, whereas electron–hole interactions that cause excitonic instability may be dominant in NaAlGe, forming a pseudogap.[39]

We have been exploring new anti-PbFCl-type compounds and novel phenomena associated with the unusual electronic structures. Other compounds with combinations of elements in groups I–III–IV include LiAlSi and LiAlGe; however, these compounds have a half-Heusler structure ($F$–$43m$).[42] There has been no report of the synthesis of NaAlSn, NaGaSi, or NaGaGe. We report on a new compound, KAlGe, in which the Na in NaAlGe is replaced by K. We successfully synthesized single crystals using the potassium-indium flux method. Electrical resistivity, heat capacity, and X-ray diffraction (XRD) measurements revealed a phase transition from one metallic phase to another below $T_s$ = 89 K, in contrast to NaAlGe's semiconducting behavior at low temperatures. The transition is accompanied by a minor structural modification that breaks the fourfold symmetry. We will examine the distinct properties of the three compounds and discuss characteristic electronic instability that underpins this group of materials, which may highlight a novel aspect in materials science.

## 2. EXPERIMENTAL

KAlGe single crystals were obtained from a melt using potassium–indium flux; NaAlSi and NaAlGe were prepared with sodium flux.[29] Adding indium to the potassium flux produced thicker crystals. The following materials were weighed at a molar ratio of K:Al:Ge:In = 3:1:1:0.5; K lump (Strem Chemicals, Inc., 2N), Al rod (5mm diameter; Nilaco, 6N), Ge lump (Kojundo Chemical Laboratory, 5N), and In (Mitsubishi Metal Corporation, 5N). The mixture was placed in a boron nitride crucible 8.5 mm in diameter and 20 mm long and sealed in a stainless-steel container 16.4 mm in diameter and 80 mm long. All operations were performed in an argon-filled glove box. The container was heated in an electric furnace to 1173 K in 3 hours, then cooled to 1073 K in 1 hour, 823 K in 80 hours, and finally furnace-cooled to room temperature. The crucible was taken out of the stainless-steel container and placed in a glass tube in the glove box. The excess potassium was removed by heating the crucible in the glass tube at 300°C for 12 hours while it was evacuated, yielding plate-like KAlGe crystals. Caution is advised because potassium can ignite when reacting with moisture and air during synthesis; to our knowledge, the only one report of single crystal synthesis using the potassium-flux method is of GaN.[43] Similar to NaAlSi and NaAlGe, this compound is unstable in the atmosphere, degrading and turning yellow after interacting with moisture for a few minutes. The chemical composition was determined by wavelength dispersive X-ray (WDX) spectroscopy using an electron probe micro-analyzer system (JEOL XA-8200).

Single crystal XRD experiments were performed using synchrotron X-rays with a wavelength of 0.69053 Å at the beamline BL-8A at the Photon Factory in the High Energy Accelerator Research Organization. Oscillation photographs of KAlGe single crystals were taken using a curved imaging plate (Rigaku R-Axis). The sample was sealed in a capillary to avoid degradation. Data integration and absorption correction were performed with Rapid-Auto software. Structural parameters were refined by the full-matrix least-squares method using the Crystal Structure program package (Olex2).[44] The temperature variation of the XRD pattern was recorded upon cooling down to 35 K.

Powder XRD patterns were obtained using Cu-K$\alpha_1$ radiation on a SmartLab diffractometer with a one-dimensional silicon tip detector. To avoid degradation, a powder sample was mixed with Apiezon N grease in a glove box and placed on a copper plate with a depth of 0.2 mm. A refrigerator was used to cool the sample to 4 K. The lattice constants were determined by the Rietveld method using the FullProf software.

The electrical resistance and heat capacity were measured using a physical property measurement system (PPMS, Quantum Design). To measure electrical resistance, the four-terminal method was used, which involved attaching a gold wire with silver paste to a crimped indium terminal on the crystal surface. Following termination in a glove box, the sample was covered with liquid paraffin to prevent degradation during the transfer from the glove box to the PPMS. Magnetic susceptibility was measured using a magnetic property measurement system (MPMS, Quantum Design). To minimize degradation, a sample pellet was greased in a glovebox and inserted into the sample chamber filled with He gas. All of the aforementioned measurements were made during slow heating following quick cooling to liquid helium temperature.

First-principles electronic structure calculations for KAlGe were carried out using the all-electron full-potential linearized augmented plane wave (FLAPW) method [45-47] implemented in the HiLAPW code [48] based on the generalized gradient approximation of the density functional theory (DFT).[49] The spin orbit interaction (SOI) was considered self-consistently for the valence and core states by the second variation scheme.[50] The energy cutoffs of wavefunction and potential expansions were 20 and 160 Ry, respectively. For the self-consistent calculations, the Brillouin zone was sampled with a tetrahedral integration scheme using a 16 × 16 × 8 mesh of $k$-points centered on Γ.[51]

## 3. Results

### 3.1. Chemical composition and crystal structure

The obtained KAlGe single crystals were bluish black and plate-like, with a habit plane of (0 0 1), widths of 1 to 4 mm, and thicknesses of 0.1 to 0.4 mm (Figure 1(b)). A typical crystal's chemical composition, as determined by WDX measurements, was $K_{0.98(1)}Al_{1.07(1)}In_{0.007(1)}Ge_{0.95(1)}$; the error represents a small scatter within a crystal, indicating uniform composition. The other two crystals shared similar compositions. Moreover, assuming the presence of indium, a crystal grown without it had the composition of $K_{0.95(1)}Al_{1.10(1)}In_{0.005(1)}Ge_{0.94(1)}$. This implies that the observed indium contamination of less than 1% could be an artifact of the experiments. The observed compositions deviate slightly from the stoichiometric composition, indicating that some K and Ge sites may have been replaced with an excess of Al. However, given the WDX analysis's empirical accuracy of a few percentage points, it is unclear whether nonstoichiometry is present in the composition.

The single crystal XRD experiments show that the crystal has an anti-PbFCl-type crystal structure (tetragonal, space group $P4/nmm$ (No.129)). Table 1 lists the crystal structure data and analysis results at 296 K. The refinement converged with small reliability factors ($R_1 = 1.68\%$ and $wR_2 = 4.45\%$). Table 2 shows the crystal structure parameters. It assumed full occupancy at all three sites; using occupancy as free parameters resulted in no improvement in refinement. The excess Al in the WDX results was most likely due to crystal surface contamination or experimental artifacts. The structural differences between NaAlSi, NaAlGe, and KAlGe will be addressed later.

**Table 1. Crystal data and results analyzed for the single crystal KAlGe.**

| Formula weight $M_r$/g mol$^{-1}$ | 138.72 |
|---|---|
| Crystal form, color | platelet, blueish black |
| Crystal system | tetragonal |
| Space group, Z | $P4/nmm$ (No. 129), 2 |
| Radiation wavelength $\lambda$/Å | 0.69053 |
| $F_{000}$ | 128 |
| Temperature $T$/K | 296(2) |
| Unit cell dimensions | |
| $a$/Å | 4.2977(4) |
| $c$/Å | 8.2708(8) |
| Unit cell volume $V$/Å$^3$ | 152.76(3) |
| Calculated density $D_{cal}$/Mg m$^{-3}$ | 1.508 |
| Limiting indices | |
| $h$ | $-6 \leq h \leq 7$ |
| $k$ | $-3 \leq k \leq 5$ |
| $l$ | $-9 \leq l \leq 13$ |
| $2\theta$ range for data collection/° | 10.7 – 72.46 |
| Data, restraints, parameters | 241, 0, 10 |
| Extinction coefficient $x$ | 0.045(8) |
| Goodness-of-fit on $F^2$; $S$ | 1.120 |
| $R_1$, $wR_2$* % | 1.68, 4.45 |
| Largest diff. peak and hole/e·Å$^{-3}$ | 1.246, –0.697 |

*$R_1 = \Sigma \ ||F_0| - |F_c|| \ / \ \Sigma \ |F_0|$,
$wR_2 = \{\{\Sigma \ w[(F_0)^2 - (F_c)^2]^2\}/ [\Sigma \ w(F_0^2)^2]\}^{1/2}$,
$w = [\sigma^2(F_0)^2 + (aP)^2 + bP]^{-1}$, where $P = [(F_0)^2 + 2(F_c)^2]/3$.

**Table 2. Atomic coordinates and equivalent isotropic displacement parameters. WP and occ. denote Wyckoff position and site occupancy, respectively.**

| atom | WP | occ. | $x$ | $y$ | $z$ | $U_{eq}$ (Å$^2$) |
|---|---|---|---|---|---|---|
| K | 2c | 1 | 1/4 | 1/4 | 0.64782(9) | 0.0192(2) |
| Al | 2a | 1 | 3/4 | 1/4 | 0 | 0.0112(2) |
| Ge | 2c | 1 | 1/4 | 1/4 | 0.18546(3) | 0.0115(1) |

3.2. Physical properties

The temperature dependence of the three compounds' in-plane electrical resistivity $\rho$ are shown in Figure 2(a). NaAlSi has an electrical resistivity of 0.9 mΩ cm at room temperature. It exhibits metallic behavior before transitioning to superconductivity at 6.8 K. NaAlGe has a resistivity of 0.3 mΩ cm at room temperature. It is semiconductive and rapidly increases in temperatures below 100 K. KAlGe has a similar electrical resistivity to NaAlGe at room temperature. However, after a sharp rise around 90 K, it resumes metallic behavior below 80 K, approaching the low-temperature value of NaAlSi. Neither NaAlGe nor KAlGe exhibits superconducting transitions above 1.8 K. The behavior of KAlGe is almost unaffected by batch growth with potassium–indium flux or indium-free potassium flux (Supporting Information 1). This suggests that nonstoichiometry and indium contamination have little, if any, effect on the properties.

Figure 2(b) shows the temperature-dependent magnetic susceptibility $\chi$ of the three compounds measured using a 7 T magnetic field in the $a$-axis direction. NaAlSi exhibits a smooth increase upon cooling, as previously reported.[35] In contrast, NaAlGe and KAlGe exhibit paramagnetic properties similar to NaAlSi at high temperatures, but their magnetic susceptibility rapidly decreases below 100 K and 90 K, respectively. These temperatures are consistent with the values at which electrical resistivity shows anomalies.

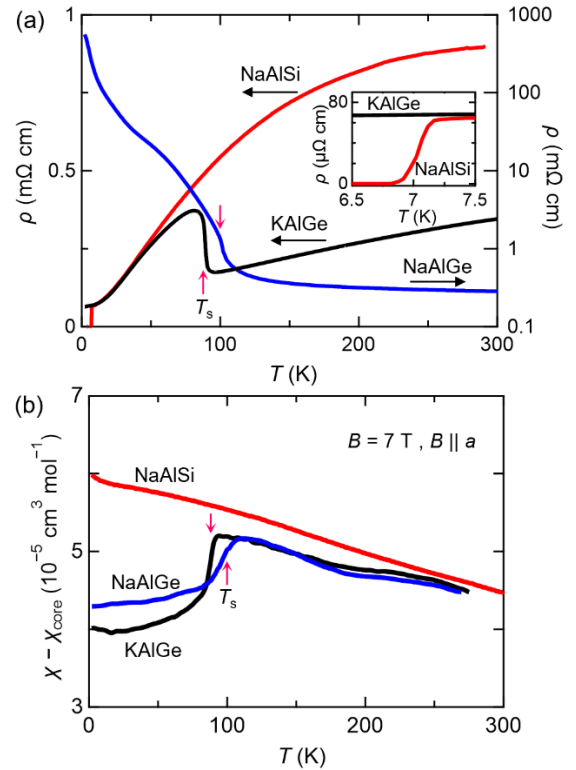

**Figure 2.** (a) Temperature-dependent in-plane electrical resistivity ($\rho$) of KAlGe, NaAlGe, and NaAlSi single crystals: data for KAlGe and NaAlSi on the left axis, and NaAlGe on the right. The inset shows an enlarged view of the low-temperature regions, where a superconducting transition occurs in NaAlSi rather than KAlGe. (b) Temperature dependence of the magnetic susceptibility of KAlGe, NaAlGe, and NaAlSi measured using a 7 T magnetic field applied along $a$ axis. The diamagnetic contribution of core electrons ($\chi_{core}$) has been extracted. Arrows in KAlGe and NaAlGe indicate $T_s$ based on heat capacity data

(Figure 3(a)). All of the measurements were made during slow heating following quick cooling to liquid helium temperature.

The temperature dependence of *C/T*, the heat capacity divided by the temperature, is shown in Figure 3(a). The values of *C/T* at 100 K are larger for NaAlSi, NaAlGe, and KAlGe, in that order, which is attributed to the decrease in Debye frequency caused by the inclusion of heavier elements. There are no anomalies in NaAlSi other than the superconducting transition at 6.8 K, while there is a clear peak around 90 K and 100 K in KAlGe and NaAlGe, respectively, indicating a phase transition. The phase transition temperature $T_s$, defined as the temperature at which the peak of heat capacity begins to grow (arrow in Figure 3(a)), is 89 K for KAlGe, which is nearly identical to the temperatures at which the electrical resistivity and other quantities such as magnetic susceptibility in Figure 2(b) or those in Figure 4 show anomalies. NaAlGe, on the other hand, exhibits a minor peak at $T_s$ = 100 K that almost corresponds to the anomalies in resistivity and magnetic susceptibility in Figure 2, indicating the presence of a phase transition similar to that observed in KAlGe. In previous research, the continuous transition from high-temperature metallic to low-temperature semiconducting behavior in NaAlGe was regarded as a crossover.[39] However, the present heat capacity data, possibly obtained from a higher-quality crystal (Supporting Information 2), show a weak phase transition.

The low-temperature heat capacity of three compounds is shown in Figure 3(b). The typical relationship $C(T) = \gamma T + \beta T^3$ almost holds below 3 K, and the Sommerfeld coefficient $\gamma$ is derived from the $C/T$ versus $T^2$ intercept. The estimated Sommerfeld coefficient value for KAlGe is 0.20(13) mJ K$^{-2}$ mol$^{-1}$, which is more than an order of magnitude smaller than NaAlSi's value of 2.166(8) mJ K$^{-2}$ mol$^{-1}$. KAlGe has a very low and finite density of states at low temperatures. The current NaAlGe crystal shows almost zero $\gamma$, with 0.07(30) mJ K$^{-2}$ mol$^{-1}$ in Figure 3(b), which is significantly smaller than the previous value of 0.45 mJ K$^{-2}$ mol$^{-1}$.[35, 40] The sample dependence is shown in Supporting Information 2. Our crystals are more insulating at low temperatures, while previous crystals' finite $\gamma$ was due to unexpected doping from chemical modifications.

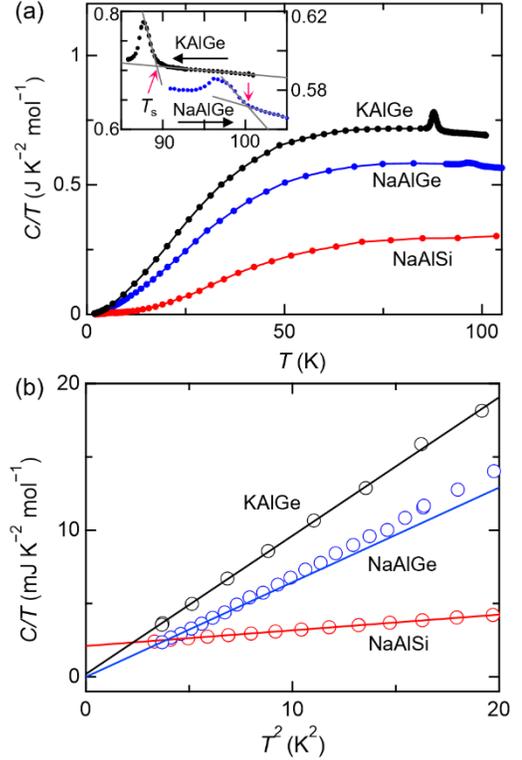

**Figure 3.** (a) Temperature dependence of the heat capacity divided by temperature (*C/T*) in KAlGe, NaAlGe, and NaAlSi single crystals. The insert shows a zoomed-in view of the temperatures where KAlGe and NaAlGe peak. The arrows point to $T_s$, the temperature at which the peak appears. (b) *C/T* versus $T^2$ plots for KAlGe, NaAlGe, and NaAlSi. The straight lines fit to the equation $C(T) = \gamma T + \beta T^3$.

The magnetic field dependence of the in-plane Hall resistance of KAlGe was linear with a negative slope, except at low temperatures when quantum oscillations overlapped (Supporting Information 3). Thus, electrons, which are relatively light, are the dominant carriers. Figure 4 shows the estimated temperature dependence of carrier density (*n*) based on the single-carrier model. The carrier density is 3.23 × 10$^{21}$ cm$^{-3}$ at room temperature and rapidly decreases at $T_s$. At 2 K, the carrier density is 1.87 ×10$^{19}$ cm$^{-3}$, over two orders of magnitude lower than the value at room temperature. The mobility ($\mu$) estimated by the equation $\mu = 1/\rho e n$ from *n*, $\rho$ and elementary charge *e* is 5.6 cm$^2$ V$^{-1}$ s$^{-1}$ at room temperature, which is an order of magnitude smaller than that of the normal metal Cu of 46 cm$^2$ V$^{-1}$ s$^{-1}$. However, it rises dramatically near $T_s$, reaching 5.2 × 10$^3$ cm$^2$ V$^{-1}$ s$^{-1}$ at 2 K, three orders of magnitude larger than the room temperature value. These findings suggest that the electronic state changes significantly at $T_s$.

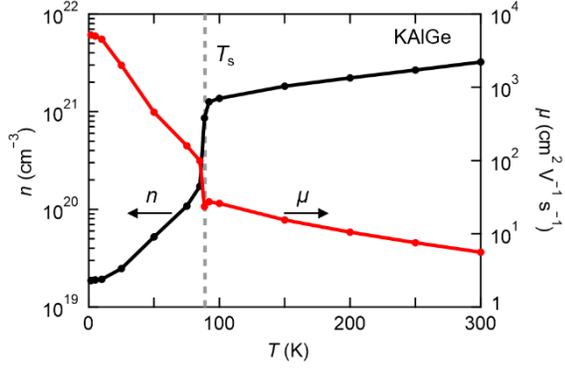

**Figure 4.** Temperature-dependent carrier density $n$ (left axis) and mobility $\mu$ (right axis) of single crystal KAlGe.

### 3.3. Structural transition

Powder XRD experiments were performed at various temperatures to investigate potential structural changes during the phase transition. The Rietveld analysis of the XRD pattern at room temperature (Figure 5(a)) yielded nearly identical structural parameters as the previous single crystal structural analysis. The 4 K XRD pattern showed no additional peaks or peak disappearance (Supporting Information 4). Moreover, there was no peak splitting, only systematic peak broadening. The 1 1 0 and 1 0 2 peaks are slightly broadened, as shown in the inset of Figure 5(a), but the 0 0 1 reflection is not (not shown). This indicates a weak structural deformation at low temperatures, with the crystal structure adopting orthorhombic or lower symmetry.

Figure 5(b) shows the temperature-dependent lattice constants derived from fitting the XRD patterns to the space groups $P4/nmm$ and $Cmme$. The $c$-lattice constant decreases smoothly as temperature decreases. At low temperatures, the $a$-lattice constant splits into orthorhombic lattice constants $a_o$ and $b_o$. According to the unit cell definitions of $P4/nmm$ and $Cmme$, $a_o$ and $b_o$ are equivalent to $a$ times $\sqrt{2}$. At the lowest temperature, the orthorhombicity $\delta$, which is defined as $\delta = (a_o - b_o) / (a_o + b_o)$, is unusually small at 0.13%. Interestingly, the splitting becomes smaller on heating, as also seen in the temperature dependence of $\delta$ in Figure S5 in Supporting Information 5. However, the splitting does not vanish at $T_s$, as expected for a conventional phase transition, but continues up to 160 K, which is much higher than $T_s$.

Although the power XRD experiments found only peak broadening, the single-crystal XRD data revealed the presence of additional superlattice reflections below $T_s$. The inset of Figure 5(c) shows the X-ray oscillation photograph recorded at 133 K and 47 K, which include the 5 1 12 and 5 1 13 reflections ($P4/nmm$). At 47 K, a superlattice reflection with the 5 1 25/2 index appears between the fundamental reflections 5 1 12 and 5 1 13. The temperature dependence of 5 1 25/2 reflection intensity in the main panel clearly shows that this peak appears at $T_s$, indicating that a structural transition indeed takes place at $T_s$. Thus, the orthorhombic distortion begins to develop at temperatures far above the structural phase transition temperature. In addition, the half-integer index reflection implies the presence of a $2c$ superlattice in the low-temperature phase. A complete structural analysis of the low-temperature phase is now in progress.

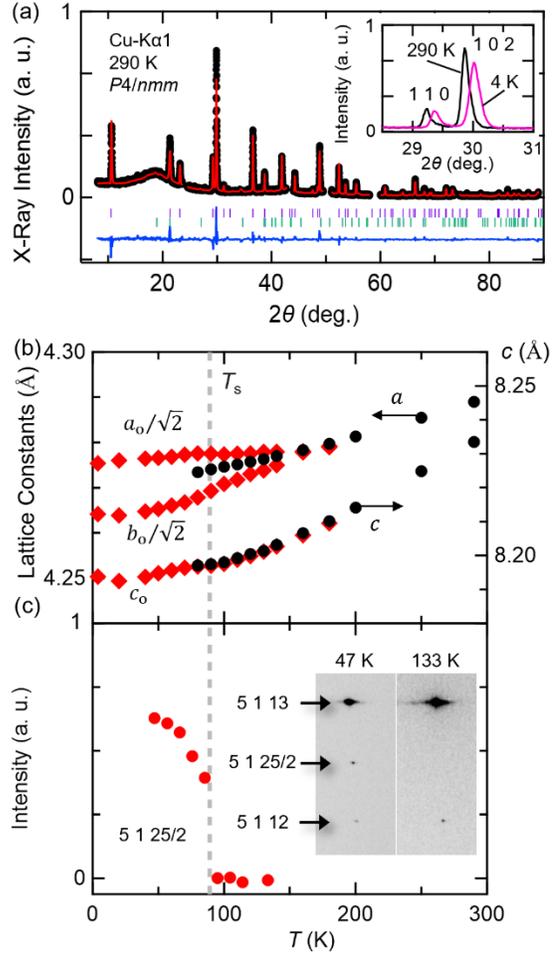

**Figure 5.** (a) Rietveld analysis of powder XRD patterns of KAlGe at 290 K. All peaks are identified with KAlGe in the $P4/nmm$ structure (purple marks) and AlOOH (green marks), the latter of which may have been produced through partial degradation. The inset shows an enlarged view of the region containing the 1 1 0 and 1 0 2 reflections, comparing the diffraction peaks at 290 K and 4 K. (b) Temperature-dependent lattice constants for the $a$ (left) and $c$ (right) axes. The values examined by $P4/nmm$ and $Cmme$ are represented as black circles and red squares, respectively. (c) Temperature-dependent intensity of the 5 1 25/2 reflection as indexed by $P4/nmm$. The inset shows the single-crystal XRD patterns for the 5 1 $l$ peak ($12 \leq l \leq 13$) at 133 K and 47 K.

### 4. Discussion

The three compounds have comparable crystal and electronic structures, but their physical properties differ significantly. NaAlSi shows no phase transitions other than superconductivity, while KAlGe and NaAlGe exhibit phase transitions of unknown origin at 89 and 100 K, respectively, but no superconductivity. The structural transition of KAlGe is characterized by a change in symmetry to orthorhombic or lower symmetry, which is accompanied by the formation of a $2c$ superlattice. Our preliminary single-crystal XRD experiments show that NaAlGe has the same structural deformation and $2c$ superlattice formation as KAlGe.[52] Thus, both Ge compounds are very similar, but their low-temperature phases differ significantly: KAlGe is a metal, whereas NaAlGe is an insulator. Here, we will look at why the ground states of these three seemingly similar compounds are so different.

We first consider it in terms of crystal structure. All three compounds form a conduction layer that consists of edge-sharing Ge/Si tetrahedra centered on an Al atom in the anti-PbFCl structure (Figure 1(a)). This conducting layer is separated by double square sheets made of K/Na ions. Table 3 lists the structural parameters for each material. In comparison to NaAlSi, the $a$-axis lengths of NaAlGe and KAlGe are enlarged by 1.0% and 4.3%, respectively, while the $c$-axis lengths increase by 0.7% and 12.3%. As a result, the structural two-dimensionality in KAlGe must be enhanced significantly. This is due to the K$^+$ ion's 38% larger ionic radius than Na$^+$. Because the $c$-axis length corresponds to the interlayer distance between the Al-Si/Ge conduction layers, KAlGe must have the most two-dimensionality in its electronic structure. On the other hand, the conduction layer has a similar shape across compounds, with little variation in the Al–Si/Ge bond length ($d$) and conduction layer thickness ($t$). Additionally, the Al–(Si/Ge)–Al bond angle ($\theta$) is consistently close to the ideal tetrahedral value of 109.5 degrees. Hence, KAlGe possesses conduction layers that are similar to those found in other compounds but are significantly separated from one another, resulting in increased two-dimensionality.

**Table 3** Comparison of structural parameters and physical properties of NaAlSi, NaAlGe, and KAlGe.

|  | NaAlSi | NaAlGe | KAlGe |
| --- | --- | --- | --- |
| $a$ [Å] | 4.1217(1) | 4.1634(2) | 4.2977(4) |
| $c$ [Å] | 7.3629(2) | 7.4146(4) | 8.2708(8) |
| $z$ (Na/K) | 0.6346(1) | 0.6355(1) | 0.6478(1) |
| $z$ (Si/Ge) | 0.20764(4) | 0.21252(2) | 0.18546(3) |
| $d$ (Al-Si/Ge) [Å] | 2.5660(2) | 2.6108(1) | 2.6402(2) |
| $\theta$ (Al-Si/Ge-Al) [deg.] | 106.86(1) | 105.75(1) | 108.96(1) |
| $t$ (layer) [Å] | 1.5288(3) | 1.5758(1) | 1.5339(2) |
| Ground state | Superconductivity | Psuedogap | Dirac like |
| Charac. temp. [K] | 6.8 | 100 | 89 |

Figure 6 compares the band structure of KAlGe (calculated using structural data at 296 K from Tables 1 and 2) to NaAlSi and NaAlGe from previous studies.[39] The semimetal state in these compounds results from the overlap of the Al-3$s$ electron band and the hole band which consists of the Si/Ge-$p_x$ and $p_y$ orbitals (Supporting Information 6). Because of the different orbital symmetry, the crossing point can be preserved against anticrossing, resulting in Dirac points in the absence of spin–orbit interactions; however, spin–orbit interactions produce small gaps that are too small to be seen in Figure 6. Among several Dirac points, the one between the Z and R points (red arrow) is closest to Fermi energy and thus has the greatest potential to influence transport properties.

The three compounds share similar band structures but differ along the Γ–Z line. KAlGe exhibits smaller band dispersion than NaAlSi and NaAlGe, indicating weaker electron transfer along the $c$ axis. This is obviously due to the enhanced structural two-dimensionality, as mentioned above. Additionally, KAlGe lacks the band that crosses the Fermi energy between Γ and Z, which is present in NaAlSi and NaAlGe. As a result, the KAlGe's Fermi surface has higher degree of two-dimensionality.

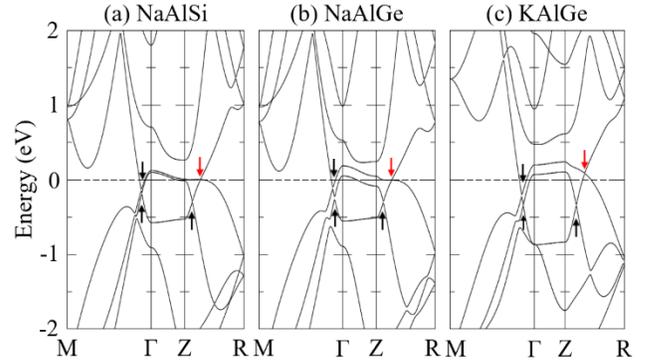

**Figure 6.** Band structures of (a) NaAlSi, (b) NaAlGe, and (c) KAlGe. The calculations dealt with spin–orbit interactions. The energy origin is set at the respective Fermi energy. The Γ, Z, M, and R points are (0, 0, 0), (0, 0, π/$c$), (π/$a$, π/$a$, 0), and (π/$a$, 0, π/$c$), respectively (see Figure S6 in Supporting Information 6). The arrows represent Dirac points at band crossings.

In terms of band structures, we discuss the origin of the phase transition at $T_s$ in both KAlGe and NaAlGe. Because the series of compounds is a $sp$-electron system, electron correlation effects are minimal, and Mott transitions are unlikely to occur. Moreover, magnetic instability has not been observed. Because KAlGe has Fermi surfaces with higher two-dimensionality, one might expect a CDW transition due to Fermi surface nesting. However, the observed 2$c$ superlattices are unrelated to nesting because they lack in-plane modulation, and no additional superlattices indicating CDW formation in the $ab$ plane are found. NaAlGe is less two-dimensional than KAlGe, making a CDW mechanism less likely. We believe that the electronic instability observed in KAlGe and NaAlGe has a common origin.

Nematic instability is a type of electronic instability that could be responsible for the phase transition in KAlGe. In iron-based superconductors, the connection between nematic instability and superconductivity has attracted a lot of attention recently.[53] The majority of iron-based superconductors undergo a phase transition from a tetragonal structure at high temperatures to an orthorhombic one at low temperatures in order to lift the in-plane degeneracy of the $d_{xz}$ and $d_{yz}$ bands. BaFe$_2$As$_2$, for instance, exhibits a phase transition to an orthorhombic structure with a small orthorhombicity $\delta$ below 130 K. Below 170 K, $\delta$ develops due to nematic fluctuations.[27, 28, 54] In KAlGe, the degeneracy between Ge-$p_x$ and $p_y$ orbital-derived bands results in a pair of saddle-shaped hole Fermi surfaces around the Γ point (see Supporting Information 6). Thus, the existence of a nematic instability in KAlGe is not surprising. Actually, it exhibits orthorhombic distortion below 160 K, which is significantly higher than $T_s$, and then increases further at $T_s$ (Supporting Information 5). KAlGe's $\delta$ value at 4 K is 0.13%, comparable to 0.45% for BaFe$_2$As$_2$[54] and 0.15% for NaFeAs[55]. Therefore, nematic instability may contribute to KAlGe. Nevertheless, whether the nematic transition can create a pseudogap remains a serious question. In KAlGe, an orthorhombic distortion is predicted to cause one of the $p_x$ and $p_y$ bands to rise while the other falls. After this operation, one of the saddle-shaped hole Fermi surfaces may disappear, but the other should remain. Therefore, this type of modification does not open the pseudogap. Furthermore, since NaAlSi with a similar band structure does not exhibit a phase transition, this instability cannot account for why a phase transition is observed in only the other two compounds.

We suggest excitonic instability as an alternative option. In NaAlSi, the instability associated with electron–phonon interactions dominant, resulting in electron–electron pair formation and superconductivity.[35] On the other hand, it has been proposed that in NaAlGe, excitonic instability causes a pseudogap to open, thereby overcoming the electron–phonon instability.[39] Excitonic instability is theoretically considered to form a pseudogap and stabilize an excitonic insulator phase, but no clear experimental evidence has been found.[21, 56, 57] The formation of exciton gaps could explain why the majority of Fermi surfaces in KAlGe and NaAlGe are removed during phase transitions. Given that KAlGe and NaAlGe have phase transitions while NaAlSi does not, Ge may be important for the excitonic instability. It is worth noting that the conventional mean-field electronic calculations do not account for excitonic instability.

Finally, we discuss the differences in the low-temperature properties between NaAlGe and KAlGe: the former has semiconducting resistivity, while the latter has metallic conductivity. In both compounds, the pseudogap opens during phase transition, causing the loss of the majority of Fermi surfaces at low temperatures. The divergent electrical resistivity of NaAlGe simply supports this picture. In contrast, KAlGe's high metallic conductivity must be due to the survival of a portion of Fermi surface with extremely light mass. The low-temperature phase of KAlGe has a mobility of $5.2 \times 10^3$ cm$^2$ V$^{-1}$ s$^{-1}$, which is comparable to the typical values of Dirac electrons reported for topological materials such as the topological insulator $Bi_2Se_3$ ($6.2 \times 10^2$ cm$^2$ V$^{-1}$ s$^{-1}$),[58] the nodal-line semimetal $CaIrO_3$ ($6.0 \times 10^3$ cm$^2$ V$^{-1}$ s$^{-1}$),[59] and the Dirac semimetal $EuMnBi_2$ ($1.4 \times 10^4$ cm$^2$ V$^{-1}$ s$^{-1}$).[60] Dirac points are found in the high-temperature phase of KAlGe (Figure 6). If some of them remain at $E_F$ after the pseudogap formation, their high mobility would be sufficient to produce metallic conductivity comparable to NaAlSi despite the low carrier density; superconductivity is not expected because of the significantly lower carrier density, or smaller density of states, than in NaAlSi. It is unclear why this happens only in KAlGe rather than NaAlGe. It could be due to a difference in two-dimensionality or a subtle difference in symmetry in the low-temperature crystal structure. KAlGe is a unique topological semimetal in which the Dirac point is "hidden" in the tetragonal high-temperature phase before manifesting itself in the lower-symmetry phase after the excitonic phase transition at low temperatures.

Further discussion requires determining the crystal structure of the low-temperature phase of KAlGe and performing band structure calculations on that structure; however, it should be noted that the Coulomb interactions that cause excitonic instability are not sufficiently taken into account in standard DFT calculations. Revisiting NaAlGe would be beneficial for gaining a systematic understanding. Angle-resolved photoemission spectroscopy or other experimental techniques should be used to determine the change in the Fermi surface during the phase transition. Although excitonic instability has been investigated in various materials, no conclusive evidence has been found, and this material system appears to be an outstanding candidate.

## 5. Conclusion

In summary, we successfully synthesized a single crystal of a new compound, KAlGe, using the potassium-indium flux method. It undergoes a metal-to-metal transition at $T_s$ = 89 K. During the phase transition, the carrier density decreases significantly while the mobility increases, resulting in high metallic conductivity in the low-temperature phase. These changes are caused by the opening of a pseudogap that contains Dirac-like electrons. At $T_s$, a structural transition from tetragonal to orthorhombic, or lower symmetry, occurs, which is accompanied by a $2c$ superlattice. The origin of the phase transition is unclear, but we believe that excitonic instability plays an important role.

The family of compounds' diverse physical properties make for intriguing future research. We believe that conventional electron–electron interactions via phonons are dominant in NaAlSi, resulting in superconductivity, whereas exotic electron–hole interactions dominate in KAlGe and NaAlGe, resulting in pseudogap formation; Dirac-like electrons survive only in KAlGe, not NaAlGe, most likely due to a subtle Fermi level difference. They offer an intriguing platform for the systematic investigation of novel electronic states.

## AUTHOR INFORMATION

**Corresponding Author**

Toshiya Ikenobe − *Institute for Solid State Physics, University of Tokyo, Kashiwa, Chiba 277-8581, Japan;*
Email: t_ikenobe@issp.u-tokyo.ac.jp

**Author Contributions**

The study was conceived and planned by T.I. All authors contributed to the manuscript, but T.I. and Z.H. wrote the majority of it. T.Y. and T.I. carried out the crystal synthesis. T.I. and J.Y. conducted powder XRD experiments, while T.I. and R.O. conducted Rietveld analysis. With assistance from H.S., R.O., and Y.O., T.I. and J.Y. conducted the majority of the single crystal XRD experiments. With assistance from D.H. for electrical resistivity methods, T.I. conducted measurements for electrical resistivity, magnetic susceptibility, heat capacity, and Hall resistance. T.O. performed first-principles electronic structure calculations. The final draft of the manuscript has been approved by all authors.

## ACKNOWLEDGMENT

The authors would like to thank Chikako Nagahama for her assistance with the sample preparation, Takashi Kamaya for assistance with the WDX measurements, Hisanori Yamane for assistance for Single crystal XRD experiment, and Daisuke Hamane for taking pictures of crystals. This research was funded by JSPS KAKENHI Grants (JP22H05147, JP22H01178, and JP23K22449), the Cooperative Research Program "Network Joint Research Center for Materials and Devices" (20235012, 20245021), JST ASPIRE (Grant Number: JPMJAP2314), and the MEXT Program "Data Creation and Utilization Type Material Research and Development" (JPMXP1122683430). The XRD study was performed with the approval of the Photon Factory Program Advisory Committee (Proposal No.2023PF-Q001).

## ABBREVIATIONS

XRD, X-ray diffraction; WDX, wavelength dispersive X-ray

## Supporting Information

### Supporting Information 1

Figure S1 shows the temperature dependence of electrical resistivity for KAlGe single crystals, comparing those prepared with potassium–indium flux in Figure 2(a) to a crystal from a different batch also prepared with potassium–indium flux, as well as a crystal prepared using indium-free potassium flux. All three crystals exhibit comparable sharp $T_s$ transitions at nearly equal temperatures, suggesting that any indium contamination and non-stoichiometry, if present, do not influence transport properties. The minor variation in resistivity magnitude at high temperatures could be due to experimental ambiguity or macroscopic crystal defects. We examined multiple additional batches of KAlGe crystals and found that they all behaved similarly.

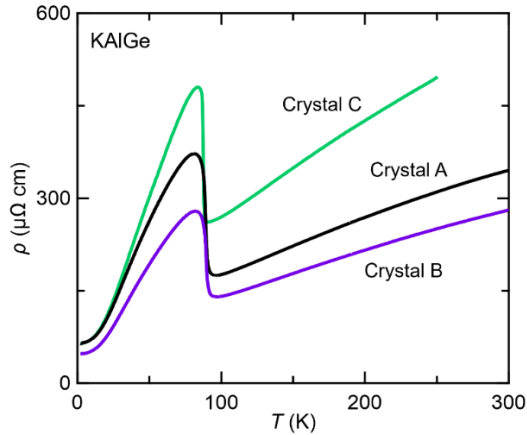

**Figure S1.** Comparison of the resistivity between the Paper's KAlGe crystal A and one more crystal B using potassium-indium flux, as well as crystal C using indium-free potassium flux.

### Supporting Information 2

Figure S2(a) shows the sample dependence of the electrical resistivity of NaAlGe. Crystal A, the crystal of this paper, exhibits more insulating behavior than the previously reported single crystal B[1] and polycrystal C[2]. Crystals D is another batch prepared in this study that shows similar insulating behavior to Crysal A, indicating little sample dependence among the present crystals. Figure S2(b) shows the corresponding sample dependence of low-temperature heat capacity. Sommerfeld coefficients for Crystal B and polycrystal C were previously reported to be 0.45(2) and 0.45(1) mJ K$^{-2}$ mol$^{-1}$, respectively. In contrast, Crystals A and E have the values of 0.07(30) and –0.16(18) mJ K$^{-2}$ mol$^{-1}$, respectively, which are practically assumed to be zero. Thus, NaAlGe has an insulating ground state with an energy gap. Previous samples' high conductivity and finite $\gamma$ were likely caused by off-stoichiometry-related doping.

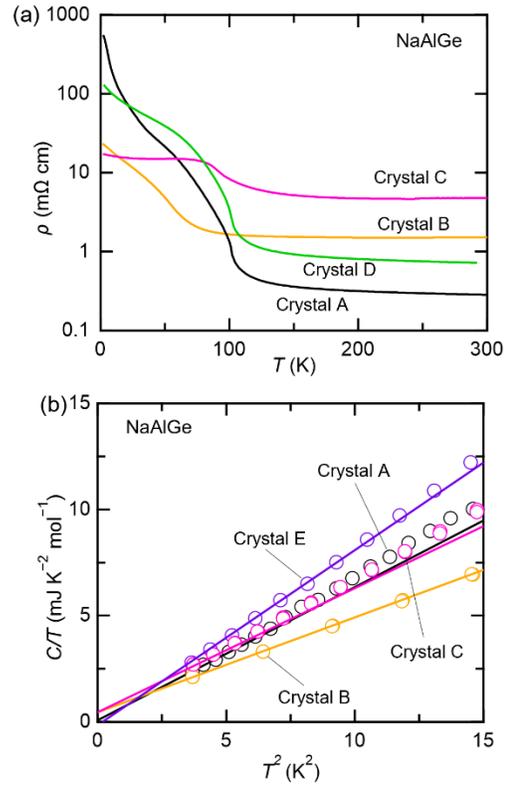

**Figure S2.** Sample dependence of (a) resistivity and (b) low-temperature heat capacity of NaAlGe. Crystal A is the single crystal used in this paper, Crystal B is a single crystal reported in reference [1], Crystal C is a polycrystal reported in reference [2], and Crystals D and E are other crystal batches prepared in this study.

### Supporting Information 3

Figure S3 shows the magnetic field dependence of in-plane Hall resistivity for KAlGe at various temperatures. It shows a linear dependence across all temperatures with a negative slope. Below 25 K, an additional oscillation component appears due to quantum oscillation, indicating that the current crystals have high crystalline quality. The electron carrier density $n$ from the slope of the linear fit to the single-carrier model, represented by $\rho_{xy} = (1/en) \times B$, and its temperature dependence is plotted in Figure 4.

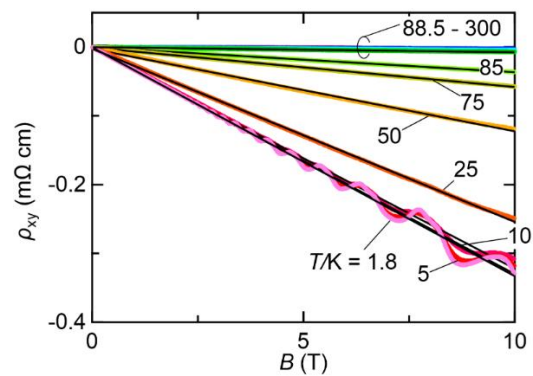

**Figure S3.** Magnetic field dependence of Hall resistance $\rho_{xy}$ for a KAlGe single crystal at different temperatures. The black line represents a linear fit to the single-carrier model.

## Supporting Information 4

Figure S4(a) shows a comparison of the powder XRD patterns of KAlGe at 290 K and 4 K. The 4 K XRD pattern exhibits neither new peaks nor the disappearance of existing ones. Figures S4(b–d) show the Rietveld analyses of the 4 K XRD pattern, which employ the high-temperature phase space group *P*4/*nmm* and its subgroups *Cmme* and *Pmmn*. The reliability factors are $R_1 = 9.76\%$ and $wR_2 = 14.6\%$ for *P*4/*nmm*; $R_1 = 6.57\%$ and $wR_2 = 10.0\%$ for *Cmme*; and $R_1 = 10.7\%$ and $wR_2 = 17.1\%$ for *Pmmn*. *Cmme*'s reliability factor has significantly improved.

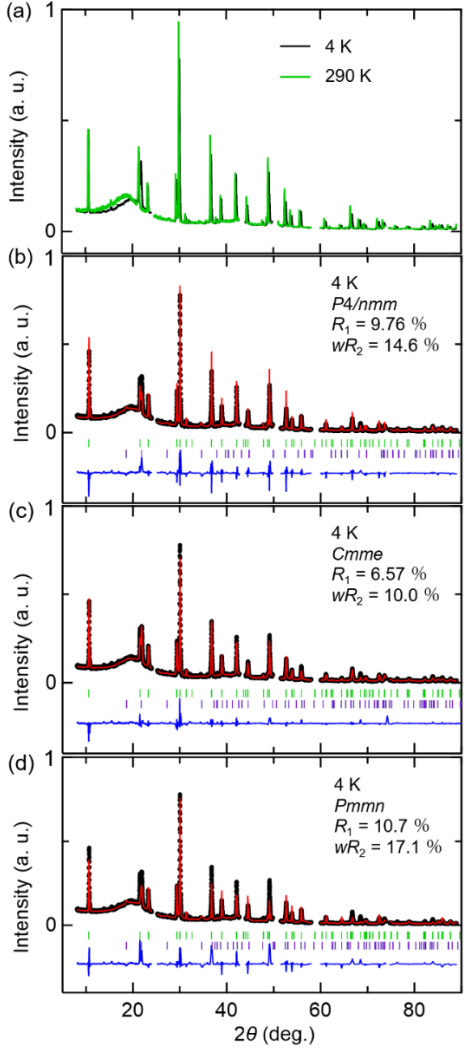

**Figure S4.** (a) Powder XRD patterns of KAlGe at 290 and 4 K. (b)–(d) Rietveld analyses of the 4 K XRD pattern with the *P*4/*nmm*, *Cmme*, and *Pmmn* models. Each fitting contains AlOOH, which could have been produced during degradation. The figures show the $R_1$ and $wR_2$ values.

## Supporting Information 5

Figure S5 shows the temperature dependence of the orthorhombic distortion $\delta$, using the lattice constants presented in Figure 5(b). The $\delta$ already appears at $T^* = 160$ K, significantly increases at $T_s$, and reaches saturation as temperature decreases further. Nematic fluctuations can raise $\delta$ beyond the phase transition temperature, as observed in other nematic material systems.[3,4]

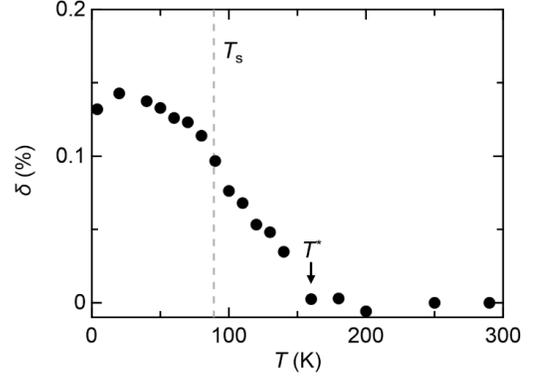

**Figure S5.** Temperature dependence of the orthorhombic lattice distortion $\delta$, calculated as $(a_o - b_o)/(a_o + b_o)$, for KAlGe.

## Supporting Information 6

Figure S6(a) displays the orbital-projected band structure of KAlGe, calculated with spin–orbit interactions, using the structural data at room temperature from Tables 2 and 3. Figure S6(b) depicts the relevant notation for positions in momentum space. The band dispersion at $k_z = 0$ is shown in Figure S6(c). Near the Γ point, two orthogonal saddle-shaped hole bands (Ge-$p_x$ and Ge-$p_y$) that are degenerate in energy intersect with the Al-$s$ electron band, resulting in the formation of Dirac points.

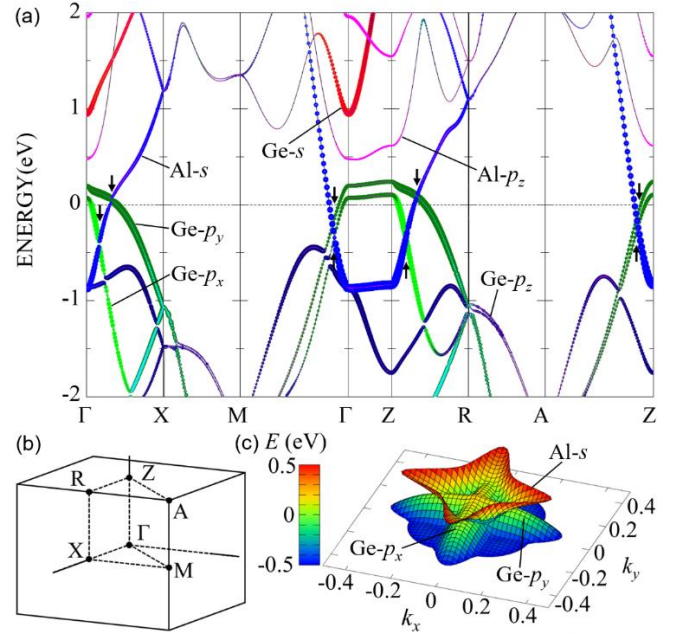

**Figure S6.** (a) Band structure of KAlGe in its high-temperature phase, calculated without spin–orbit interactions. Each orbit is distinguished by color, while the circle's size represents the state's weight. The arrows depicted in the figure represent band crossings. (b) Brillouin zone in a tetragonal crystal structure. (c) Band dispersion at $k_z = 0$.